\documentclass[aps]{revtex4}
\usepackage{makeidx}
\usepackage{bm, amsmath, amssymb}
\usepackage{graphicx,subfigure}
\usepackage{amsfonts}
\usepackage{amsmath}
\usepackage{amssymb}
\usepackage{graphicx}
\usepackage[cp1251]{inputenc}
\usepackage[T2A]{fontenc}
\usepackage{color}
\newcommand{\cl}{\textcolor{black}}
\newcommand{\as}[1]{\textcolor{black}{#1}}

\begin{document}
\title{Nonlinear spin dynamics of ferromagnetic ring in the vortex state and its application for spin-transfer nano-oscillator}

\author{Vera Uzunova}
\affiliation{Institute of Theoretical Physics, Faculty of Physics, University of Warsaw, ul. Pasteura 5, 02-093 Warszawa, Poland}
\affiliation{Institute of Physics of the National Academy of Sciences of Ukraine, 46 Nauky Ave., 03039 Kyiv, Ukraine}

\author{Boris A. Ivanov}\email{bor.a.ivanov@gmail.com}
\affiliation{Institute of Magnetism, National Academy of Sciences of Ukraine, 03142 Kiev, Ukraine}
\affiliation{Institute for Molecules and Materials, Radboud University Nijmegen, Heyendaalseweg 135   6525 AJ Nijmegen The Netherlands}

\date{\today}

\begin{abstract}
We study a nonlinear spin dynamics of a ferromagnetic ring in a vortex state induced by the spin-polarized current.
We also suggest to use the ferromagnetic ring as a free layer of a coreless vortex spin-transfer nano-oscillator. The calculated working frequency is about several GHz, that is much higher than the gyromode frequency of the disk-based vortex oscillator. The response of the vortex-state ring to the spin-polarized current has hysteretic behavior with the reasonable values of the thresholds current densities: ignition threshold is about $10^{8} \text{A}\text{cm}^{-2}$, and elimination current to maintain the oscillations has much lower values about $10^{6} \text{A} \text{cm}^{-2}$. The output signal can be extracted by the help of the inverse spin Hall effect or by the giant magnetoresistance. The output electromotive force averaged over all sample vanishes, and we suggest to use a ferromagnetic ring or disk in a vortex state as a GMR analyzer. For an inverse spin Hall analyser we advise to use two heavy metals with different signs of Spin-Hall angle. The ring-based STNO is supposed to increase the areas of practical application of the STNOs.
\end{abstract}

 \maketitle

\section{Introduction}

 Spin-polarized current flowing into a magnetic material creates a large torque acting on the magnetization due to the direct transfer of spin angular momentum, that opens up opportunities to manipulate mesoscopic magnetic elements \cite{Slonczewski}. Discovery of a spin transfer mechanism laid the foundation of a new generation of spintronic devices \cite{Kiselev, BA7, Ralph, Bader},  including auto-oscillators \cite{Ruiz-Calaforra,ISHE1,ISHE2,Houssameddine}, magnetic random
access memory \cite{Vaysset} and logic gates \cite{Murugesh,BA5}.

One of the key spintronic elements applicable in wireless communications technology are spin-transfer nano-oscillators (STNOs), \cite{GMR2}. These devices are using the spin transfer torque to generate microwave signals by transforming energy from a dc electric current into high-frequency magnetic oscillations. Traditionally designed STNOs are fabricated as sandwich structures of magnetic and nonmagnetic nanolayers in forms of nanocontacts \cite{Consoloa,Soucaille,Houshang} or nanopillars, \cite{Tarequzzaman,Zeng,GuslienkoAranda}.  The main structural elements of the STNO are a free layer, i. e. a thin magnetic layer allowing rotation of the magnetization, a polarizer and an analyzer.  Polarizer, the thicker magnetic layer with relatively fixed magnetization, gives a spin polarisation to direct current passing through the device. Above a certain critical current density its spin-transfer torque acting on a thinner free magnetic layer can locally overcome the intrinsic damping and excites a steady-state spin auto-oscillations on one or more spin-waves modes of the system.  In the analyser these spin oscillations are converted to a microwave power by the magnetoresistive effect or by the Inverce Spin Hall  effect.

 Advantages of the STNOs are small size, broad range of working temperatures and easy integration into standard semiconductor technology. A characteristic feature of all STNOs is nonlinear dynamics of the free layer magnetization.  Typical current densities for creating a steady-state spin dynamics are $10^7 -10^8 \text{A} \text{cm}^{-2}$ in nanocontacts and $10^6 -10^7 \text{A} \text{cm}^{-2}$  in nanopillars \cite{Zeng1}.

A promising alternative to the single-domain oscillators are vortex STNOs based on ferromagnetic nano-disks in a so-called vortex state. The vortex configuration has a closed magnetic flux and creates demagnetising fields only in a small region of the vortex core.  Therefore magnetic vortex can realize the ground state of magnetic nanoparticles. The vortex STNOs are characterized by a uniquely narrow line width,
require lower current densities and dissipate less energy comparing to single-domain oscillators \cite{Zvezdin2009, Zaspel_07}.  They are useful for applications in arrays of oscillators, memory systems on vortex states, and others. The primary excitable mode in the vortex-state particle is the gyrotropic mode (gyromode) that corresponds to the slow precessional motion of the vortex around the disc center. The frequency of the gyromode is usually lower then 1 GHz. This slow dynamics of the vortex core is used in the standard design of the vortex-state STNO. Operating frequencies of the vortex oscillators are low compared to single-domain oscillators, where frequencies of $10$-$15$ GHz can be achieved.

The generated frequency can be doubled by using of a dual-free-layer STNO because of superposition of two opposite current-driven precessional motions of vortices in two free layers, \cite{Prokopenko}.  One more way to increase the working frequency of the vortex-state STNOs is to use ferrimagnets near the point of compensation of the angular momentum. They have gyrotropic frequencies much higher than those of ferromagnets  \cite{Tserkovnyak, ZaspelGalkinaIvanov19} that allows the operating frequency up to $20$ GHz. Limitations of this design are related to the larger critical size of the vortex state ferrimagnetic disk with the thicknesses less than 10 nm, which is needed to spin-torque applications. Therefore, the problem how to rise the working frequency of the vortex-state STNO is quite challenging.

In this paper we investigate nonlinear spin dynamics of a ferromagnetic ring in a vortex state under an action of spin-polarized current and discuss the usage of the ferromagnetic ring as the base of a vortex-state STNO. Ferromagnetic nanorings are widely investigated due to their unique optical and electromagnetic properties, \cite{BA3};   biomedical applications of vortex-state magnetic rings \cite{BA1,BA2} are one of the current interests. The dynamics of magnetization in the vortex-state ring differs significantly from that for the vortex-state disk, \cite{ZaspelJMM}. The frequency of the lowest magnon mode for the ring is much higher then frequency of the gyromode of vortex-state magnetic disc, that opens up new opportunities to rise the working frequency of the STNO. The ring-based vortex oscillator is expected to have some significant advantages comparing to the case of oscillators based on vortex-state discs, providing high values of the working frequencies, and smaller size. 

The adequate theoretical description of the non-linear spin dynamics for the vortex-state ring is much more complex then for the particles with uniform (or quasi-uniform) magnetization, which can be analysed within the macrospin approximation. In our work we used the collective variables approach, which is valid for highly-nonlinear regimes of oscillations for the wide variety of magnetic solitons, see e.g. \cite{Oleg}. We found the inertial behavior of the spin dynamics of the vortex-state ferromagnetic ring, formally common to one for antiferromagnetic STNO, with corresponding analogy of oscillators based on superconducting Josephson junction \cite{SCR}. One of the important consequences is the lower value of the critical current, the elimination critical current, which is needed to support the generation once it has been started. Thus the ring-based STNO can have suitable geometric dimensions and threshold current densities and therefore is a good candidate for practical applications.

The article is organized as the following. In Section \ref{model} basic equations describing magnetization dynamics in the system are given. The ground state of the ferromagnetic ring and spin dynamics above this ground state are considered in Sections \ref{gs} and \ref{breath}, respectively. In Section \ref{jcr} we obtain expressions for critical currents and working frequency. Finally, in Section \ref{signal} we discus ways of extracting a useful ac electric output signal for the case of interest when the output electrical power averaged over the contact is zero. 

\section{ Ferromagnetic ring under the action of the spin-polarized current. \label{model}}
\begin{figure}[h]
\includegraphics[width = 0.32\linewidth]{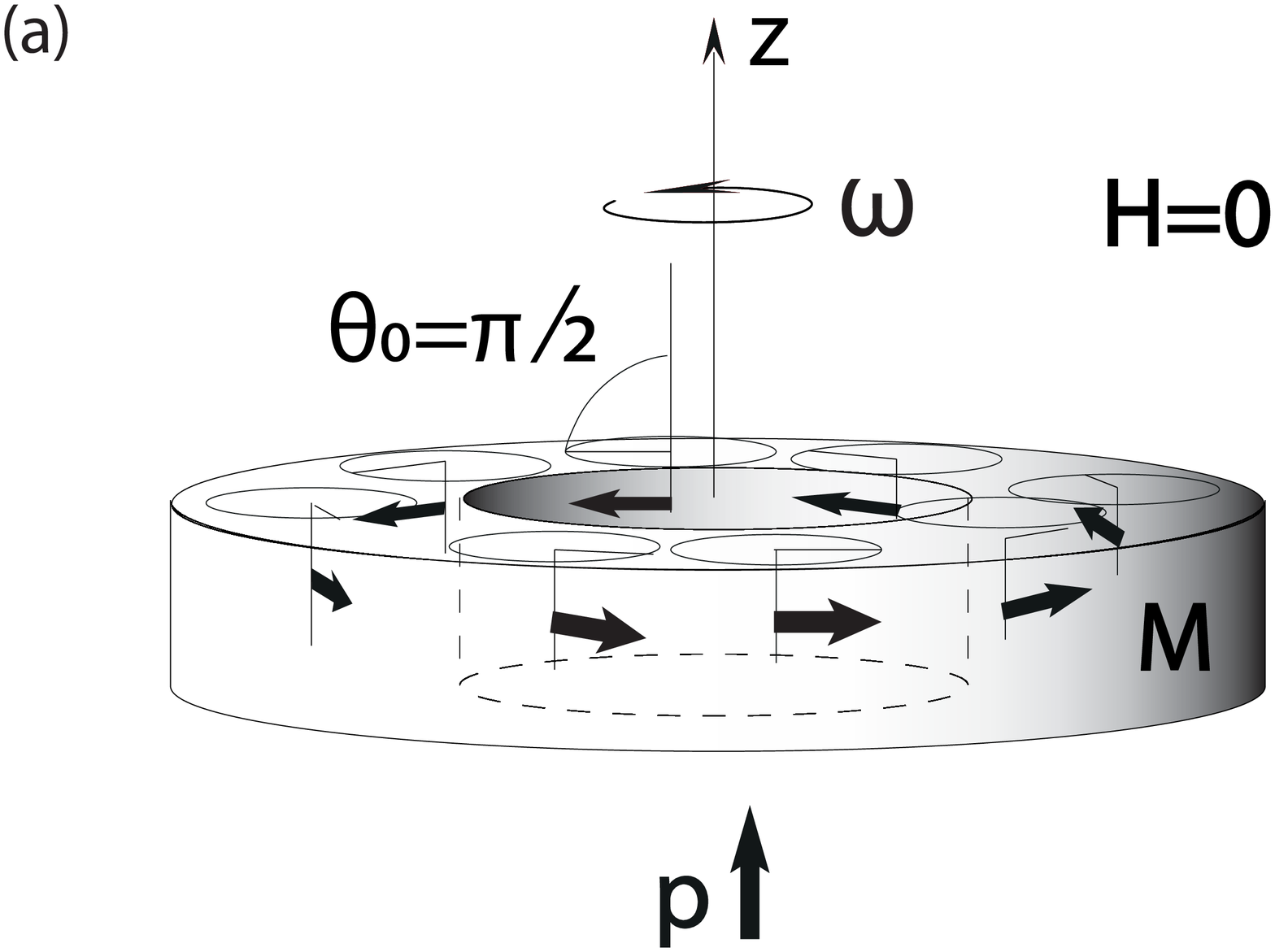}
\includegraphics[width = 0.30\linewidth]{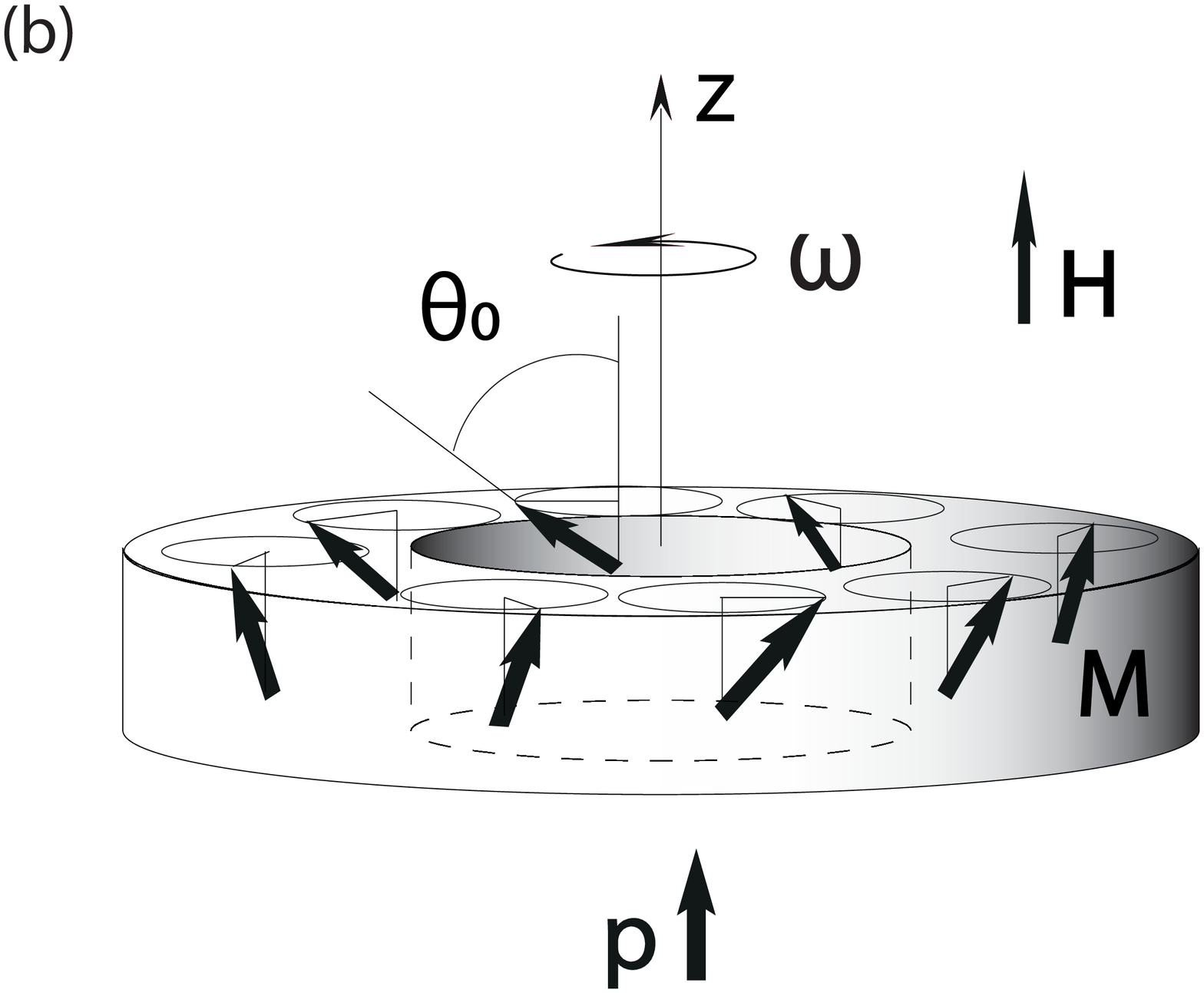}
\caption{Ground state and dynamics of magnetization $\mathbf{M}$ of the vortex-state ring under the action of spin-transfer torque: (a) standard in-plane vortex with the polar angle $\theta_0=\pi/2$. (b) cone-vortex state, which realizes  the ground state of the free layer when the external magnetic field $\mathbf{H}$ is applied along the $z$-axis, the angle $\theta_0 \neq \pi/2$. The spin-polarized current  with polarization $\mathbf{p}$ flows into the free layer. Vector $\boldsymbol{\omega}$ shows schematically the direction of rotation of magnetic moments in different points of the ring.\label{Fig1}}
\end{figure}

The problem under consideration is to investigate magnetization dynamics induced by the spin-polarized current in a thin ferromagnetic ring. Spatial dimensions of the ring are: thickness $L$, inner and outer radii $R_i$ and $R_o$ respectively; ring is considered thin, $L\ll R_{i,o}$. Ground state of the ring is the vortex state, in which magnetization $\mathbf{M}$ aligns circularly around the ring axis. It is discussed in section~\ref{gs}. In the Cartesian reference frame with the $z$-axis along the ring axis the components of the ground-state magnetization are $M_s(-\sin\theta_0\sin\chi,\sin\theta_0\cos\chi,\cos\theta_0)$, where $\chi$ is the azimuthal coordinate in the plane of the ring and  $\theta_0$ is the polar angle of the ground-state magnetization. The schematic picture of the vortex-state ring and its magnetization are shown in Fig.~\ref{Fig1}. 

The spin-polarized current flows into the ring and its spin-polarization $\mathbf{p}$ is along $z$. The current flow is assumed to be uniform within the sample.  Spin-polarized current creates a torque $\mathbf{T}$, which tends to rotate the in-plane component of $\mathbf{M}$ in the plane of the ring. 

In addition we are considering an option of applying a weak (with the value less than saturated) magnetic field $\mathbf{H}$ along $z$-axis: $h=H/4\pi M_s$, where $M_s$ is the saturation magnetization of the ferromagnetic material of the ring. If the external magnetic field is zero, $h=0$, magnetization $\mathbf{M}$ lies entirely in the plane of the ring and the polar angle $\theta_0=\pi/2$ , Fig.~\ref{Fig1}(a). The case of nonzero $\mathbf{H}$ is shown in Fig.~\ref{Fig1}(b). The small external field, $h< 1$,  leads to appearance of a small out-of-plane component of magnetization equal to $M_s h$. Then the polar angle $\theta_0$ in the ground state is determined by the external field as $\cos \theta_0=h$. We  show below that small external field does not change the nature of the excitations but may give an additional way of the device control.

The dynamics of magnetization $\mathbf{M}$ is described by the Landau-Lifshitz equation 
\begin{eqnarray}\label{M1}
\frac{\partial\mathbf{M}}{\partial t}=\gamma\left[\mathbf{M}\times\frac{\delta
W}{\delta\mathbf{M}}\right]+\frac{\alpha}{M_s}\left[\mathbf{M}\times\frac{\partial\mathbf{M}}{\partial
t}\right]+\mathbf{T}.
\end{eqnarray}
Here $W\equiv W[\mathbf{M}]$ is the ferromagnet's energy written as a functional of the magnetization $\mathbf{M}$, the vector $-\delta
W/\delta\mathbf{M}$ has the sense of the effective field, $\gamma = g \mu_B/\hbar $ is the gyromagnetic ratio, $g$ is Lande factor, $\mu_B$ is the Bohr magneton, $\hbar $ is the Planck's constant, $\alpha$ is the dimensionless parameter of the Gilbert damping. The particular form of the energy functional $W$  will be presented in the next section. 
The spin-transfer torque is given by $\mathbf{T}=\sigma J
M_s\left[\mathbf{m}\times[\mathbf{m}\times\mathbf{p}]\right]$, where
$\mathbf{m}=\mathbf{M}/M_s$ is the unit magnetization vector of the free layer, $\mathbf{p}$ is the spin-polarization of the current. The value of $\sigma$ is given by the formula

\begin{eqnarray}\label{}
\sigma=\frac{\epsilon g \mu_B}{2 e M_s L S },
\end{eqnarray}
where $e$ is the modulus of the electron charge, $\epsilon$ is dimensionless spin-polarization efficiency, and $S$ is the area of the current-carrying region that in our case coincides with the area of the ring face.

Considering the spin dynamics of the ring it is convenient to parameterize the vector of normalized magnetization $\mathbf{m}$ by two angles: azimuthal angle $\varphi$  and polar angle $\theta$. Then the Cartesian components of $\mathbf{m}$ are $(\sin\theta\cos\varphi,\sin\theta\sin\varphi,\cos\theta)$. The energy functional $W$ can be also presented as a functional of these angles $W=W[\theta,\varphi]$. Then the dynamic equations take the form

\begin{eqnarray}\label{eq1}
\frac{M_s}{\gamma}\sin\theta\frac{\partial\varphi}{\partial
t}=\frac{\delta W}{\delta\theta}+\frac{\alpha
M_s}{\gamma}\frac{\partial\theta}{\partial t},\\\label{eq2}
-\frac{M_s}{\gamma}\sin\theta\frac{\partial\theta}{\partial
t}=\frac{\delta W}{\delta\varphi}+\frac{\alpha
M_s}{\gamma}\sin^2\theta\frac{\partial\varphi}{\partial
t}-\frac{\sigma J M_s\sin^2\theta}{\gamma}.
\end{eqnarray}

\as{
These equations can be presented as the Lagrange equations for a non-conservative system 
\begin{eqnarray}\label{Lc}
\frac{\partial \mathcal{L}}{\partial \theta}-\frac{\partial}{\partial t}\frac{\partial \mathcal{L}}{\partial (d\theta/dt)}=\frac{\partial Q}{\partial (d\theta/dt)},\notag \\
\frac{\partial \mathcal{L}}{\partial \varphi}-\frac{\partial}{\partial t}\frac{\partial \mathcal{L}}{\partial (d\varphi/dt)}=\frac{\partial Q}{\partial (d\varphi/dt)},
\end{eqnarray}
with a dissipative function $Q$ and
a Lagrangian density
\begin{eqnarray}\label{LL}
{\mathcal{L}}=\frac{M_s L}{\gamma}\int\left[ \cos\theta
\frac{\partial\varphi}{\partial
t} -W[\theta,\varphi]\right]d^2\mathbf{r}.
\end{eqnarray}
\cl{
For a thin magnetic particle, the problem is essentially two-dimensional and the integration is performed over two-dimensional vector $\mathbf{r}$, that lies in the plane of the ring with an origin at its center. The integration over $z$, where $z$-axis is perpendicular to the ring plane, simply results in factor $L$.}
}

\as{The dissipative function of the ferromagnet is $Q=Q_\text{G}+Q_\text{STT}$, where 
\begin{eqnarray}\label{QG}
Q_\text{G}=\frac{\alpha M_s L}{2\gamma}\int\left[ \left(
\frac{\partial\theta}{\partial
t}\right)^2+\sin^2\theta\left(\frac{\partial\varphi}{\partial
t}\right)^2\right]d^2\mathbf{r},
\end{eqnarray}
\begin{eqnarray}\label{QSTT}
Q_\text{STT}=-\frac{\sigma J M_s
L}{\gamma}\int\sin^2\theta\left(\frac{\partial\varphi}{\partial
t}\right)d^2\mathbf{r}.
\end{eqnarray}
Here $Q_\text{G}$ is the Gilbert dissipative function, leading to the friction force linear in velocity, as in Eqs.~(\ref{eq1}, \ref{eq2}), whereas the second term $Q_\text{STT}$ describes the contribution of the spin-transfer torque. 
The value $Q_\text{G}>0$ has a definite sign, while the sign of $Q_\text{STT}$ depends on the relative orientation of the spin-current polarization $\bm{p}$ and the magnetization $\bm{M}$. It can play the role of positive and negative friction, which determines the possibility of "anti-damping" effects. 
}


\as{The full time derivative of the energy can be presented as ${dW}/{dt}=-2Q_\text{G}-Q_\text{STT}$.} At a certain value of current $J$, these two terms can be compensated on average, giving $dW/dt=0$. It means that a regime of a steady-state oscillations can exist in the system. Our goal is to investigate possible oscillations of the system and to find conditions necessary for this regime to appear. In the following two sections  we consider in details the ground state of the ring and its excited states.


\section{Vortex state of the ferromagnetic ring \label{gs}}

Let us briefly discuss the ground state of the ring in the vortex state.  The energy of soft ferromagnet like permalloy contains the isotropic exchange interaction, the magnetostatic interaction and the interaction with the external magnetic field:
\begin{eqnarray}\label{W}
 W[\mathbf{M}]=\int\left(\frac{A}{2 M_s^2}\sum_i{\partial_i\bm{M}}^2-\frac{1}{2}\bm{M}\cdot
 \bm{H}_m-\bm{M}\cdot
 \bm{H}\right)d\bm{r}.
\end{eqnarray}
\cl{
Here $A$ is the exchange constant and 
${\bm{H}}=4\pi M_s h \bm{e}_z$ is the external magnetic field, normal to the plane of the ring, $\bm{e}_z$ is the unit vector along $z$-axis.
The field $\bm{H}_m$  is determined by the magnetostatic equations $\textmd{div} (\bm{H}_m + 4\pi
\bm{M}) = 0$ and $\textmd{rot} \bm{H}_m = 0$ with the standard boundary conditions: the continuity of the normal component of $(\bm{H}_m + 4\pi \bm{M})$ and the tangential component of $\bm{H}_m$ on the border of the sample.}
\cl{These equations can be rewritten as $\textmd{div} \bm{H}_m = 4 \pi \rho_M$, where $\rho_M $ is the ``density of magnetic charges'', which include volume charges $-\textmd{div}\bf{M}$ and surface charges, which equal to $-(\bm{M} \cdot\bm{n})$, where $\bm{n}$ is the unit vector normal to the border.
}

\cl{If the distribution of magnetization within the particle is known, what is the main assumption of the collective coordinate approach used below, the magnetostatic field $\bm{H}_m$ can be presented as a sum of the volume contribution
$\boldsymbol{H}_\textmd{m}^{\textmd{vol}}$ and surface contributions of the lateral surfaces
 ${\bm{H}}_\textmd{m}^{\textmd{edge}}$ and of the ring faces $\boldsymbol{H}_\textmd{m}^{\textmd{face}}$.
 This representation was used for description of both vortex ground state and magnon modes on the vortex background, and is proved to give a good agreement with the experimental data for vortex-state discs and rings, see \cite{Zaspel_05,ZaspelJMM}.}

\cl{For the given form of the function $\bm{M}(\mathbf{r})$, the magnetostatic energy can be present through the known $\bm{H}_m$ as the sum of corresponding volume and surface contributions. For the case of interest, the thin magnetic ring, the main term arising from the surface charges of the two ring faces is
${\bm{H}}_\textmd{m}^{\textmd{face}}=-4\pi (\bm{M}\cdot \bm{e}_z)$ that leads to the corresponding energy density of the form of $2\pi (\bm{M}\cdot \bm{e}_z)^2$.  
} 

The energy functional $W[\mathbf{M}]$ in terms of angles $\theta$ and $\varphi$ is
\begin{eqnarray}\label{en}
W[\theta,\varphi]=L\int
\left\{\frac{A}{2}
\left[(\nabla\theta)^2+
\sin^2\theta(\nabla\varphi)^2\right]+2\pi M_s^2\cos^2\theta-H M_s\cos\theta\right\}
d^2\mathbf{r} +W_{vol}+W_{edge}. 
\end{eqnarray}
First term in Eq.~(\ref{en}) represents the exchange energy with exchange constant $A$. The second term is the approximate contribution to the magnetostatic energy from upper and lower faces of the ring; this term can be treated as an effective easy-plane anisotropy, see e. g., \cite{Zaspel05}. This term is the largest of the magnetostatic terms in the case of a thin particle. The third term in the integrand is the Zeeman energy $-(\mathbf{M}\cdot \mathbf{H})=-4\pi h M_s^2 \cos \theta $ of the interaction with external magnetic field (if there is any) applied along the $z$-axis. Terms $W_{vol}$ and $W_{edge}$ denote the parts of the magnetostatic \cl{energy caused by volume and edge ``magnetic charge densities'', respectively. At the moment we do not specify the form of these contributions to the magnetostatic energy.}

The ground state magnetization distribution in the ferromagnetic ring is the coreless magnetic vortex. Introducing planar coordinates $(r,\chi)$ in the plane of the ring and $z$-axis along the axis of the ring, we can write it as $\varphi=\chi+\pi/2$ and $\theta=\theta_0(r)$, see Fig.~\ref{Fig1}. The function $\theta_0(r)$ is a result of minimization of the energy functional Eq.~(\ref{en}). Here it is taken into account that volume and edge "magnetic charge densities"  for both vortex-type configurations,  in-plane and cone, do not contribute in the energy. So, the ground state distribution of $\theta_0(r)$ is described by the solution of ordinary differential equation
\begin{eqnarray}\label{theta_eq}
l_0^2\left(\frac{d^2\theta_0}{dr^2}+\frac{1}{r}\frac{d\theta_0}{dr}-\frac{1}{r^2}\sin\theta_0\cos\theta_0\right)+\sin\theta_0\cos\theta_0-h\sin\theta_0=0,
\end{eqnarray}
where $l_0^2=A/4\pi M_s^2$ is the exchange length, and $h=H/4\pi M_s$ is the normalized external magnetic field, see for details \cite{Guslienko, Otani}.

This equation describes vortex solutions for the out-of-plane projection of magnetization $M_\perp$; it is applicable to both disk-shaped and ring-shaped particles. For a zero external field, $h=0$, the ground state corresponds to the in-plane vortex shown in Fig.~\ref{Fig1}(a), whereas for non-zero external field smaller then saturation value, $h<1$, the cone vortex state shown in Fig.~\ref{Fig1}(b) is realized, \cite{Wysin_02}. Eq.~(\ref{theta_eq}) has a singularity at $r\rightarrow 0$ that leads to appearance of vortex core (small region with out-of plane magnetization) for the ferromagnetic disk. However, for the ring the singular region of ferromagnet is just ``cutted off'' and the dynamics of excitations substantially changes. Size of the vortex core is about $l_0/\sqrt{1-h^2}$ and in what follows we assume that the field is not very close to saturation such that the core size do not exceed the inner radius of the ring. Therefore, for estimates in the main approximation in small parameters $l_0/R_o$ and $l_0/R_i$ asymptotic of Eq.~(\ref{theta_eq}) can be used: the out-of-plane component of magnetization far from the origin is $M_\perp=M_s\cos\theta_0=M_sh(1+l_0^2/r^2)$, implying $\theta_0(r)=\pi/2$ in case $h=0$. Note, that the magnetic field applied along $z$ axis reduces the out-of-plane component of the magnetization and, as a consequence, suppresses the effect of the spin-polarized current.

\section{Radially-symmetric nonlinear dynamics \label{breath}}
Excitation spectra of soft-magnetic nanoparticles are of interest both from a fundamental point of view and in context of their applications in spintronic devices. If such a particle is used as a working layer of STNO, the lower part of its spectrum determines the operating frequencies of the device. The linear dynamics of magnetic particles in the vortex state is well studied. Excitation modes in linear approximation are denoted by two integers: the number of radial nodes in the out-of-plane component of the dynamical magnetization, radial mode number $n\geq 1$, and the azimuthal mode number $m$, which determines the angular dependence of this component. The spectrum of the vortex-state disk includes the gyrotropic mode ($n=1, \ m=1$) with low (sub-gigahertz) frequency, corresponding to the slow vortex core precession around the disc center, and a system of higher modes. 

Contrary, in the spectrum of the vortex-state ring there is no low-frequency gyroscopic mode, since the vortex core is absent for the ring-shaped sample, \cite{ZaspelJMM}. At the same time, the higher mods for discs and rings are similar.

In the ring-shaped magnetic particle the mode with the lowest frequency is the breathing mode with the azimuthal number $m=0$. It is characterized by the deviation of magnetization from the ground state of the form $\varphi=\chi+\pi/2+\psi(\mathbf{r}, t)$, $\theta=\theta(\mathbf{r}, t)$, with the small deviation $\psi$. The spin waves are described by the set of dynamic Landau-Lifshitz equations. There, Eqs.~(\ref{eq1},\ref{eq2}), the ferromagnetic energy $W(\theta,\varphi)$ contains, among other terms, the nonlocal contribution to the magnetostatic energy caused by the volume and edge magnetic charges, which greatly complicates the analytical description. However, in the linear approximation this problem can be solved for small values of parameter $L/R_{o,i}$, see e.g., \cite{Zaspel05, Zaspel_05, Sheka2004}. 

The challenge of our problem is that the spin dynamics excited by the spin-polarized current is essentially nonlinear and approaches suitable for small excitations do not work. For this reason, we use the collective variable approach, derived in works \cite{ZaspelIvanov09, Galkin} \as{to describe the nonlinear analogue of the breathing mode as well as the total in-plane rotation of the magnetization vector at each point of the ring. The last one is the case of our spesial interest, as this regime can potentially provide the generation of microwave signal}.

Within the collective variable approach applied to the case under consideration the Lagrangian of the ferromagnet can be written in the form
\begin{eqnarray}\label{L1}
\mathcal{L}=\frac{L}{\gamma}\int M_\perp\frac{d\varphi}{dt}d^2 \mathbf{r}-W[\mathbf{M}],
\end{eqnarray}
where it is taken into account that magnetization $\mathbf{M}$ depends only on the in-plane coordinates. We assume that $\varphi$ is coordinate-independent variable, and its time derivative ${d\varphi}/{dt}$ can be taken out of the integral, see for details  \cite{ZaspelIvanov09, Galkin}. With this assumption the breathing mode can be described as excitations with radial symmetry: $\varphi=\chi+\pi/2+\psi(t)$, $\theta=\theta({r},t)$. Here $\psi(t)$ is the averaged value of $\psi(\mathbf{r}, t)$, the deviation from the ground state configuration. Then the Lagrangian describing the radially-symmetric dynamics of the vortex-state ring can be written through collective variables as
\begin{eqnarray}\label{L2}
\mathcal{L}=\frac{1}{\gamma}\mu\frac{d\psi}{dt}-W(\mu,\psi),
\end{eqnarray}
where $\psi(t)$ has the meaning of generalized coordinate corresponding to the azimuthal rotation of the magnetization. The quantity ${\mu}/{\gamma}$ has the sense of generalized momentum conjugated to $\psi(t)$; it is expressed through the total out-of-plane magnetic moment of the ring, ${\mu}$, namely
\begin{eqnarray}\label{mu}
\frac{\mu}{\gamma}=\frac{L}{\gamma}\int M_\perp d^2\mathbf{r}=\frac{LM_s}{\gamma}\int\cos\theta d^2\mathbf{r}.
\end{eqnarray}

In accordance with the collective coordinate approach, the Hamilton function $W(\mu,\psi)$ represents the ferromagnet's energy functional $W[ \mathbf{M}]$ minimized over the distribution of magnetization $\mathbf{M}$ at fixed values of generalized variables $\mu$ and $\psi$. This dependencies of $W(\mu,\psi)$ need to be found in the main approximation, considering the ratios $l_0/R_{i,o}$ to be small parameters.

The dependence of the energy $W$ on the generalized momentum ${\mu}/{\gamma}$ can be understood by considering how the external magnetic field applied along the axis of the ring increases the out-of-plane component of the magnetization $M_\perp$ and transforms the in-plane magnetic vortex into the cone-state vortex.
The minimum of the energy $W(\mu,\psi)$ at a given value of $\mu$ can be obtained by the method of indefinite Lagrange multipliers, see \cite{Galkin}. As a result, the extremals of the energy functional correspond to the solutions of the Eq.~(\ref{theta_eq}), where the role of magnetic field $h$ is played by the Lagrangian multiplier $h(\mu)$. Thus, apart from the vortex core region we can assume that $\cos\theta=h$ . The relation between $\mu$ and $h$ is described by Eq.~(\ref{mu}) and in the first order approximation has the form $\mu=M_sLSh$, where $S=\pi(R_0^2-R_i^2)$ is the area of the ring face. 

To obtain the explicit form of the $\mu$-dependence of the energy, we multiply Eq.~(\ref{theta_eq}) for the cone state by $r^2 d\theta/dr$, integrate it over $r$, and substitute in the energy functional Eq.~(\ref{en}), see \cite{Voronov}. Finally,
\begin{eqnarray}\label{Wmu}
W(\mu)=4\pi M_s h\mu + 4\pi^2 M_s^2 l_0^2 L (1-h^2)\ln\left(\frac{R_{o}}{R_{i}}\sqrt{\pi \Lambda (h)}\right)-SL(2\pi M_s^2 h^2),
\end{eqnarray}
where the function $\Lambda(h)\approx 5.27$ as $h\rightarrow 0$ and linearly decreases to zero at $h\rightarrow 1$, see \cite{ZaspelIvanov09}.

The $\psi$-dependence of the ferromagnet energy $W$ is determined by nonlocal effect of edge $(\mathbf{r}\cdot\mathbf{M})/r=M_s\sin\theta\sin\psi$ and volume  $(\nabla\cdot\mathbf{M})=[(1/r)d(r\sin\theta)/dr]\cos(\varphi-\chi)$ magnetostatic charges, that can be estimated by asymptotic evaluation of the magnetostatic integral, \cite{ZaspelIvanov09,Zaspel_05}. In the leading (logarithmic) order we can write
\begin{eqnarray}\label{Wedge}
W(\psi)=2\pi M_s^2L^2 \left[\left.R_o\sin^2\theta_0\right|_{r=R_o}\ln\left(\frac{\eta R_o}{L}\right)+\left.R_i\sin^2\theta_0\right|_{r=R_i}\ln\left(\frac{\eta R_i}{L}\right)\right]\sin^2\psi.
\end{eqnarray}

This expression includes the energy of the outer, $r=R_o$, and inner, $r=R_i$, edges of the ring. The contribution of the volume charge can be taken into account by a phenomenological coefficient $\eta$, it has values between $4$ and $6$, see. \cite{Zaspel05}. For the ring with $R_{o,i}\gg L$  the value of $\sin\theta_0$ on the outer and inner edges can be replaced by their asymptotic value $\sqrt{1-h^2}$.

Variables $\mu$ and $\psi$ can be considered as conjugated coordinate and momentum in accordance with Lagrangian Eqs.~(\ref{L1},\ref{L2})  and the corresponding equations for them have the Hamiltonian form. Taking into account the explicit dependence of $W$ on $\psi$, equations for collective variables $\mu$ and $\psi$ can be written as

\begin{eqnarray}
-\frac{1}{\gamma}\frac{\partial \mu}{\partial t}=\left.\frac{\partial W}{\partial \psi}\right|_{\mu=const}=\label{h1}\\
=4\pi (1-h^2)M_s^2L^2\left[ R_o\ln\left(\frac{\eta R_o}{L}\right)+R_i\ln\left(\frac{\eta R_i}{L}\right)\right]\sin\psi\cos\psi,\notag\\
\frac{1}{\gamma}\frac{\partial \psi}{\partial t}=\left.\frac{\partial W}{\partial \mu}\right|_{\psi=const}.\label{h2}
\end{eqnarray}

To go further, note that the in-plane deviation of magnetization from the ground state  weakly changes the ferromagnet energy, which makes possible nonlinear radial oscillations of the system. On the contrary, the deviation of the out-of-plane component $M_\perp$ strongly changes the vortex energy. This fact allows to separate in the main approximation the energy contributions from the azimuthal and out-of-plane deviations of magnetization. Since $\mu$ is a more rigid variable that does not contain any small parameter, it is sufficient to take into account  $\mu$  in the linear approximation even for description of highly nonlinear dynamics with unlimited change of $\psi$.  So, in the linear approximation on $\mu-\mu_0$, the derivative $\partial W/ \partial\mu$ can be replaced by $(\mu-\mu_0)(d^2W(\mu)/d\mu^2)|_{\mu=\mu_0}$, where $\mu_0$ the is equilibrium value of the out-of-plane component of the magnetic moment in the vortex core.

Then the dynamic equations take the form of the Hamilton equations for coordinate $\psi$ and conjugated momentum $p=(\mu-\mu_0)/\gamma$. Within this approximation, we can  significantly simplify the Hamilton function of our problem 
\begin{eqnarray}
H=\frac{p^2}{2m}+W(\psi),\;\;\frac{1}{m}=\gamma^2\left.\frac{d^2W(\mu)}{d\mu^2}\right|_{\mu=\mu_0},
\end{eqnarray}
where the effective mass $m$ does not depend on $\psi$ and $p$, and the potential energy $W(\psi)$ does not depend on $p$. The form of the Hamiltonian is the same as for a massive particle subject to potential energy $W(\psi)$. To the lowest order the effective mass can be evaluated using $dW(\mu)/d\mu=h(\mu)$ and the first
order approximation relation between $h$ and $\mu$,
\begin{eqnarray}\label{m}
\frac{1}{m}\approx\frac{4\gamma^2}{L(R_0^2-R_i^2)}.
\end{eqnarray}

The second Hamilton Eq.~(\ref{h2}) takes the form $\mu=\mu_0+m\gamma (d\psi/dt)$.
Taking the time derivative of it and eliminating $d\mu/dt$ from Eq.~(\ref{h1}) we come to the second-order differential equation for coordinate $\psi(t)$ in the form
\begin{eqnarray}\label{eq0}
\frac{d^2\psi}{d
t^2}+\omega_0^2(1-h^2)\sin\psi\cos\psi=0,
\end{eqnarray}
where
\begin{eqnarray}\label{}
\omega_0=\omega_m\sqrt{R_o\ln\left(\eta\frac{R_o}{L}\right)+R_i\ln\left(\eta\frac{R_i}{L}\right)}\sqrt{\frac{L}{\pi(R_0^2-R_i^2)}},
\end{eqnarray}
and $\omega_m=4\pi\gamma M_s$ is a characteristic frequency of the ferromagnetic material (of the order of 30 GHz for permalloy).  It can be seen that the frequency of $\omega_0$ contains the small parameters $L/R_{o,i}$, and is much lower than $\omega_m$. Note, that in the linear approximation, the frequency is $\sqrt{1-h^2}\omega_0$  decreases not only with decrease of relative thickness of the ring, but also with the application of the external field $h$.

Eq.~(\ref{eq0}) has an integral of motion $E=K+U$ that can be considered as a sum of kinetic and potential parts,
\begin{eqnarray}\label{E}
E=\frac{1}{2\omega_m}\left(\frac{d\psi}{dt}\right)^2+(1-h^2)\frac{\omega_0^2}{2\omega_m}\sin^2\psi.
\end{eqnarray}
The first term can be treated as a kinetic energy and the second term as a periodic potential of the restoring force of a magnetostatic nature. Thus, within the above collective coordinate approach, the typical ``inertial'' features, common to ones that are known for usual mechanics of massive particles, are found here for in-plane magnetization dynamics in the vortex-state ferromagnetic ring. Before going further, it worth discussing these features.  

The inertial dynamics is widely accepted for antiferromagnets, whereas the spin dynamics of ferromagnets is commonly believed to be gyroscopic. In particular, the equation similar to Eq.~(\ref{eq0}) was obtained for antiferromagnetic STNO and was claimed as the clear manifestation of inertial character of spin dynamics \cite{SCR}. For antiferromagnets, the inertial dynamics appears naturally within the sigma-model approach; it is based on the presence of the uniform exchange interaction between antiferromagnetic sublattices. It is valid up to extremely high frequencies of the order of the exchange frequency values, exceeding usual magnon frequencies for antiferromagnets. Inertial dynamics for some models of ferromagnets is known for a long time for some particular cases, see e.g., \cite{Mikeska1978,Kulagin}; it was also employed for description of STNO based on easy-plane magnets in uniform state \cite{inertia2022}. The models allowing such simplification are characterized by the presence of effective anisotropy with significantly different constants: hard for one angular variable and weak for the other one. In fact, it is the case of our problem. Then the inertial approach is valid up to frequency corresponding to "hard" anisotropy field. In our case, this upper limit frequency is $\omega_m$; it is much lower than the exchange frequency, but much higher than the value of interest, $\omega_0$. Thus, the inertial approximation is valid for our problem. 

\begin{figure}[h]
\includegraphics[width = 0.6\linewidth]{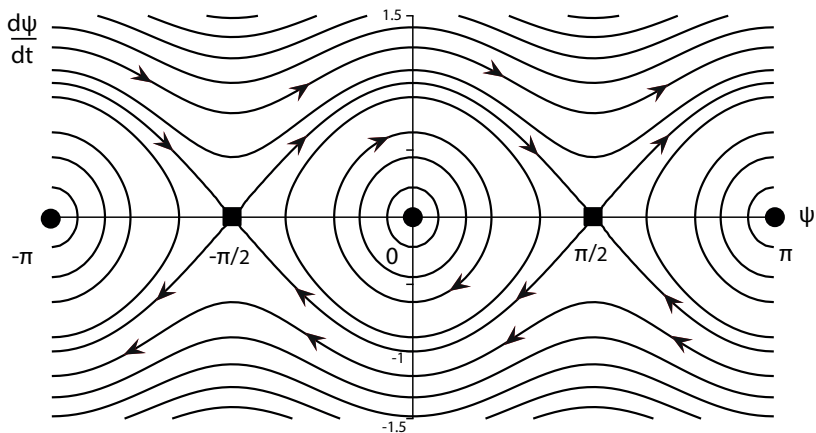}
\caption{Phase plane representation of radially-symmetric oscillations without dissipation. Here we used the energy units of $2\omega_m/\omega_0^2(1-h^2)$, and time units of $\omega_0\sqrt{1-h^2}$. Black circles show the singular points of the center type and black rectangles show saddle points. Separatrix trajectories connect the saddle points. Closed phase trajectories below the separatrix correspond to the finite motion, whereas phase trajectories above the separatrix correspond to infinite dynamics with non-limited growing of $\psi$. \label{Fig2}}
\end{figure}

The inertial approximation leads to the simple and transparent physical picture of the nonlinear motion of magnetization. Using the first integral \eqref{E}, it is possible to write down solutions of Eq.~(\ref{eq0}) in quadratures and to present them explicitly through the elliptic integrals. However, it is useful to carry out a qualitative analysis using the phase plane, the phase diagram is shown in Fig.~\ref{Fig2}.  

Stationary points of the function $E$ define singular points of the phase plane, $d\psi/dt=0$ and $\sin 2\psi=0$, and the sign of $\cos 2\psi$ determines types of these points.  There are two types of singular points: centers, $\psi=0,\pm\pi$ and $d\psi/dt=0$, around which small oscillations can occur with the frequency $\sqrt{1-h^2}\omega_0$; and saddle points, $\psi=\pm\pi/2$ and $d\psi/dt=0$. Saddle points are characterized by the constant $E=\omega_0^2/2\omega_m$, and separatrices follow the equation
\begin{eqnarray}\label{dpsi}
\frac{d\psi}{dt}=\pm\omega_0 \sqrt{1-h^2}\cos\psi.
\end{eqnarray}

This phase diagram shows that there are two regimes of oscillations: the regime with the finite oscillations of $\psi$ around the equilibrium position and the regime with unlimited increase in $\psi$ with time. These two regimes are separated by separatrices passing through the saddle points. As we will show below, these two regimes are essentially different when a ferromagnetic ring is used as an active element of the nanooscillator. In the finite regime the time average of $d\psi/dt$ equals to zero and the spin current cannot excite this type of oscillations. We are interested in the case when the initial vortex ground state of the ring loses its stability under the action of a spin-polarized current and the spin system goes into the regime of infinite circular oscillations. It takes place when the current overcome some critical values that we are about to estimate in the next section.

\section{Non-conservative dynamics and critical currents\label{jcr} }

\as{The non-conservative effects of the spin-polarized current and damping are taking into account by use of the general form of the dissipative function $Q=Q_\text{G}+Q_\text{STT}$ with Eqs.~(\ref{QG},\ref{QSTT}) and explicit form of the radially-symmetric excitation, $\theta=\theta_0$ and $\varphi=\chi+\pi/2+\psi(t)$. 
Substituting it into the Lagrangian density, Eq.~(\ref{LL}),
and using the Lagrange equation for non-conservative system, Eqs.~(\ref{Lc}),
we obtain the wave equation for $\psi$ in the Newtonian form}  
\begin{eqnarray}\label{diff_eq}
\frac{d^2\psi}{d
t^2}+\omega^2_0(1-h^2)\sin\psi\cos\psi+\omega_m(1-h^2)\left(\alpha\frac{d\psi}{dt}-\sigma
J\right)=0.
\end{eqnarray}

This equation describes a wide variety of physically interesting systems. In particular, the same equation describes the planar dynamics of the Neel vector for biaxial antiferromagnet \cite{SCR}. Note as well a simple mechanical analogy: after the substitution $\psi = \phi/2$, Eq.~(\ref{diff_eq}) coincides with the equation of a physical pendulum, a massive suspended particle moving in the gravitational field with accounting for viscous friction force (the term with $\alpha$ here) and subject to some eddy force, which is perpendicular to the radius vector of the material point that mimicries the action of spin current, see e.g., \cite{Kent, Ivanov2020}. Probably the most interesting analogy is that the above equation, written for the variable $\phi = 2\psi $, also coincides with the equation for the phase $\phi$ of the superconducting order parameter of the point Josephson contact, see for details \cite{Josephson1, Josephson2}. 

From the Eq.~\eqref{diff_eq} it follows the law of the energy evolution
\begin{eqnarray}\label{DeltaW}
\frac{d E}{dt}=\frac{M_sLS}{\gamma}(1-h^2)\left[\sigma J-\alpha\frac{d\psi}{dt}\right]\frac{d\psi}{dt}, 
\end{eqnarray}
which can be rewritten in the form $d\tilde{E}/dt=-\alpha (d\psi/dt)^2$, where $\tilde{E}=K+\tilde{U}$ and
\begin{eqnarray}\label{}
\tilde{U}=(1-h^2)\frac{\omega_0^2}{2\omega_m}\sin^2\psi-\sigma J(1-h^2)\psi.
\end{eqnarray}

Function $\tilde{U}$ includes the contribution of the spin-polarized current, which turns it into a ``tilted washboard''. So the eddy force in Eq.~\eqref{diff_eq}, that is proportional to $\sigma J$, can be formally presented as a part in the  ``potential'' $\tilde{U}$. Then its non-potential nature manifests itself in the fact, that function $\tilde{U}$ is changing after a full rotation of magnetization, i.e., after  the angle $\psi$ changes by $2\pi$. 

First note, that the system becomes absolutely unstable when local minima and maxima of potential $\tilde{U}(\psi)$ disappear, i.e., $d \tilde{U}/d\psi$ never equal to zero. Then the magnetization vector rotates through the full angle for any initial conditions. To achieve this regime, the current $J$ should overcome a critical value that is natural to call the ignition threshold
\begin{eqnarray}\label{Jc1}
J_{cr1}=\frac{\omega_0^2}{2\omega_m\sigma}.
\end{eqnarray}

\begin{figure}[h]
\includegraphics[width = 0.6\linewidth]{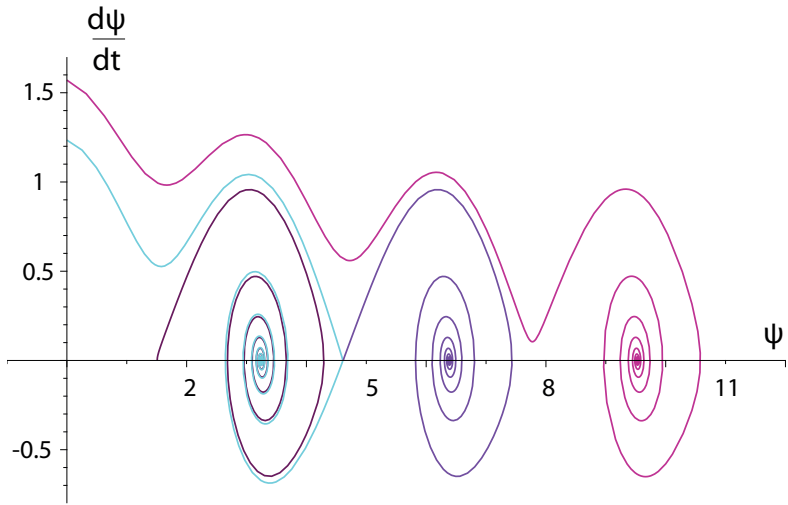}
\caption{Phase plane representation of the wave equation with the spin-polarized current $J$ and damping for the case $J<J_{cr2}$. All trajectories are tend to the focuses. We also show some of the trajectories, passing as close as possible to the separatrix trajectories from above and below. \label{FigPD1}}
\end{figure}

\begin{figure}[h]
\includegraphics[width = 0.6\linewidth]{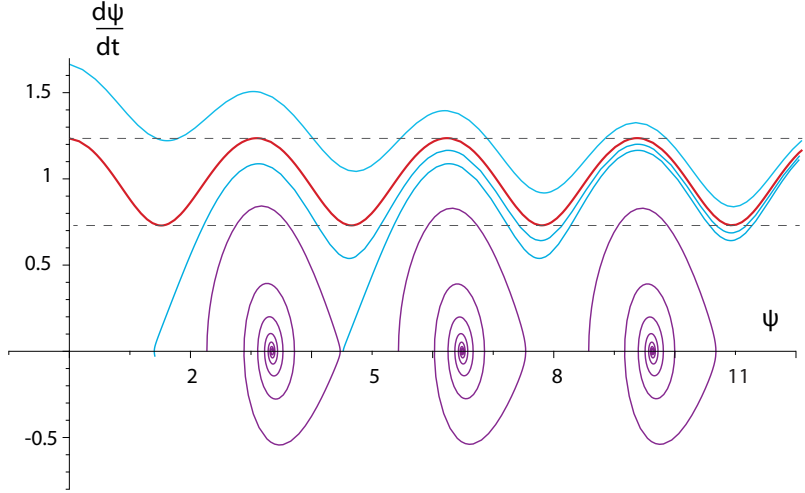}
\caption{Phase plane representation of the wave equation with the spin-polarized $J$ current and damping for the case ignition. All trajectories below the separatrix are finite. All trajectories above the separatrix are approaching the limit cycle. The limit cycle (red, color online) is shown between the two horizontal dashed lines. \label{FigPD2}}
\end{figure}

If the current value is less than $J_{cr1}$, the behavior is more complicated, but its full qualitative analysis can be done through the phase plane method. To illustrate it, some characteristic phase trajectories obtained from the numerical solution of the Eq.~(\ref{diff_eq}) for various initial conditions and a fixed current value $J$ are shown in Figs.~\ref{FigPD1},\ref{FigPD2}. 

For non-conservative system, the maxima of the effective potential still are saddle points, whereas the centers for the weak enough dissipation transform to focuses, which, in principle, could be either stable or unstable (spiral sinks or spiral sources for the phase trajectories, respectively). 

The steady-state motion and antidamping features are possible only for the infinite motion of $\psi$, i.e., for the full rotation of the magnetization. Thus, the minima of the effective potential corresponds to the stable focuses, and the phase trajectories approach them with spiralling as it shown on the figures~\ref{FigPD1} and \ref{FigPD2} below. Obviously, for zero current it is the fate of the separatrix trajectories, which coming out from the saddle points, and it is clear that the same behavior will be present at small enough values of the current. This behavior is shown in Fig.~\ref{FigPD1}. All trajectories, regardless of the initial conditions, are twisted around one of the focuses, and all oscillations in the system will decay in time. 

For the oscillations to become steady-state the change of energy of the system over a period of rotation has to be zero. This behavior appears starting from some current value, such that the phase trajectories coming out from one saddle point come into the adjacent saddle point. This condition is held for all pairs of the neighboring saddle points, and the separatrix has the same form as for the system without dissipation and pumping, see Fig.~\ref{Fig2} above.  

The corresponding critical value of the current can be found from the condition of compensation of the damping and spin pumping on this separatrix trajectory. This brings us to a concept of the second critical current, the so-called elimination threshold, $J_{cr2}<J_{cr1}$. 

Here and below we consider small damping regime, $\alpha \ll 1$, to obtain the analytical formulae for the characteristics of oscillations. This condition allows us to estimate the integrals over trajectories with use of non-perturbed equations found through Eq.~\eqref{E}, see the previous Section. All quantities are obtained in the linear approximation over the small parameter $\alpha $.

The value $J_{cr2}$ is found by substituting the equation for the separatrix trajectory, Eq.~(\ref{dpsi}) into Eq.~(\ref{DeltaW}) and integrating it over time. Finally,
\begin{eqnarray}\label{Jc2}
J_{cr2}=\sqrt{1-h^2}\frac{2\alpha\omega_0}{\pi\sigma}.
\end{eqnarray}

When the current value is above $J_{cr2}$, still some of the trajectories are finite, tending to the focuses on the phase plane. They correspond to the damped oscillations. But the trajectory coming out from one saddle point is going above the neighboring saddle point, and later on, these trajectories are going higher and higher, see Fig.~\ref{FigPD2}. 
On the other hand, for the trajectories in the region of the phase plane with high enough ``velocity'' the dissipation is more efficient, and in the average they are going ``downstairs''. Such a behavior is typical to the system with limit circle: these two sets of trajectories can be fitted to each other if and only if we suppose that all of them are tending, from both above and below, to the some periodic trajectory, limit cycle, which corresponds to the steady-state oscillations. The center line of the limit cycle is horizontal showing that in this regime the action of spin current and damping are compensated at average.

It is worth to stress here the difference of two regimes of oscillations. The ignition current is strong enough to remove the potential barriers and the system unavoidably begin to roll down the "washboard". The elimination current gives the system enough energy to inertially overcome the potential barriers of potential $U$ like a pendulum uses its kinetic energy to maintain the oscillations.

\as{We are interested in the regime of the steady-state oscillations, characterized by the full rotation, that can be used to produce a useful microwave signal. To trigger this regime, the ignition current higher than $J_{cr1}$ need to be applied for a short time $\Delta t$. The required time can be estimated from the angle needed for a complete turn rotation as $(d\Psi / dt)\Delta t=2\pi$, that gives $\Delta t=2\pi\alpha/(\sigma J_{cr1})$.}

\as{ The method of the phase plane is a good tool for understanding the dynamical behavior of the system and it allows to find the conditions for different oscillation regimes. The magnetization dynamics can be easily imagined, if we notice, that the axes on the phase diagram have the following meaning. The axis $\Psi$ is the instantaneous value of deviation of magnetization from the ground state and $d \Psi/ dt$ is its instantaneous velocity.}

\as{Magnetization distribution is shown in Fig.~\ref{Fig1} for two different ground states. 
 The dynamics of the breathing mode is the simultaneous oscillations of the magnetization vector at each point of the ring around the ground state position. For the linear breathing mode the maximal deviation from the ground state $\Psi$ is small, for nonlinear breathing mode it is up to the angle of $2\pi$. These oscillations are in the plane of the ring, and the direction of the initial deviation is symbolically shown by circled arrow denoted with $\boldsymbol{\omega}$. Spiral or closed phase trajectories on the phase diagram correspond to periodic or damped oscillations of the breathing mode with the maximal deviation smaller then $2\pi$. The wavy phase trajectories reaching the limit cycle, the regime of the steady-state oscillations, correspond to the permanent full in-plane rotation over the angle $2\pi$ of the magnetization vector at each point of the ring in the direction of $\boldsymbol{\omega}$.}

\as{The starting points of the phase trajectories in Fig.~\ref{FigPD1} and \ref{FigPD2} play the role of initial conditions, but, in fact, we do not need to specify them to achieve one of the oscillation regimes. Since the phase trajectories never intersect, it is enough to get into one of the points of the phase plane leading to the trajectory of the corresponding type. In practice the initial velocity, i.e the initial value of $d \Psi/ dt$, can be easily controlled by the magnitude of the ignition current, applied to trigger the oscillations.
The ignition current above $J_{cr1}$ is enough to reach the  regime of steady-state oscillations.
}

\begin{figure}[h]
\includegraphics[width = 0.6\linewidth]{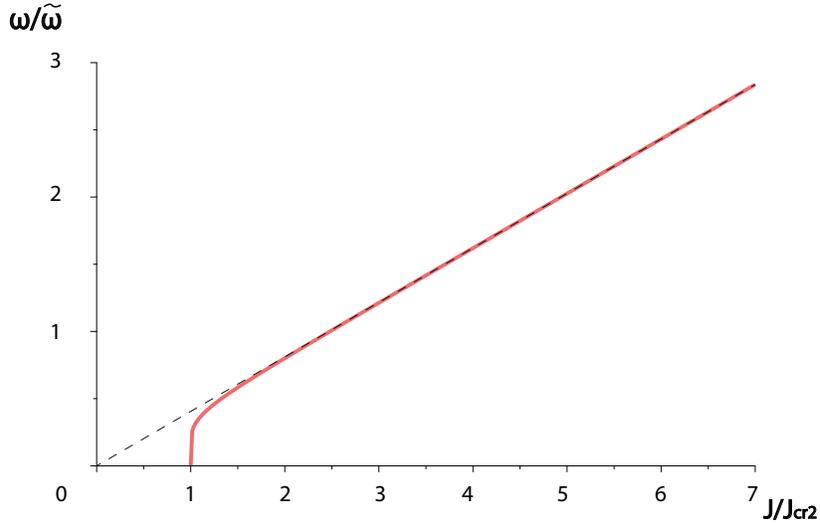}
\caption{The generation frequency $\omega$ via the value of the spin-polarized current $J$ (red line, color online). Black dashed line shows the linear asimptotic $\omega_{lin}$. The frequency units are chosen as $\tilde{\omega}=\omega_0\sqrt{1-h^2}\pi/2$. \label{Om_gen}}
\end{figure}

Let us estimate the frequency of the steady-state oscillation, corresponding to the limit cycle on the phase diagram. It will play the role of the generation frequency $\omega$ of the STNO based on the ring-shaped vortex-state free layer. In the approximation $\alpha\ll 1$ the generation frequency is found by integrating of Eq.~(\ref{E}) over the period, that gives
 \begin{eqnarray}\label{omega}
\omega=\frac{\pi}{2}\frac{\omega_0\sqrt{1-h^2}}{ k K(k)},
\end{eqnarray}
where $K(k)$ is the elliptic integral of the first kind. To find the value of argument $k$, corresponding to the limit cycle, we use the condition that at the limit cycle the integral over the period of $dE/dt$ is zero. The integrating of Eq.~(\ref{DeltaW}) gives the equation $\sigma J\pi k =2\alpha\sqrt{1-h^2}\omega_0 E(k)$, where $E(k)$ is the elliptic integral of the second kind. These two equations give the dependence of $\omega$ on the current in the implicit form, this dependence is shown in Fig.~\ref{Om_gen}.
\as{
Expressions for elliptic integrals we are using are the following
 \begin{eqnarray}\label{}
K(k)=\int_0^{\pi/2}\frac{d\beta}{\sqrt{1-k^2\sin^2\beta}},
\end{eqnarray}
 \begin{eqnarray}\label{}
E(k)=\int_0^{\pi/2}{d\beta}{\sqrt{1-k^2\sin^2\beta}}.
\end{eqnarray}
}
 
It turns out, that at current values 2-3 times higher than the threshold value $J_{cr2}$, the frequency grows almost linearly with the current, $\omega_{lin}=\sigma J/\alpha$. For small current values, $\omega$ is more complicated function of the current; it vanishes at $J \to J_{cr2}$. Despite the fact that the dependence $\omega(J)$ present in Fig.~\ref{Om_gen}  is derived in the small-$\alpha$ approximation,   $\omega_{lin}$ coincides with the linear frequency expected at the extremely high current values. The reason is, that at current values much larger than $J_{cr1}$ the role of the potential $U$ is negligible, and the limit cycle is presented by almost strait horizontal line, which corresponds to the constant solution of Eq.~(\ref{diff_eq}), namely  $d\psi/dt = \omega_{lin}$ .

To evaluate practical benefits of the ring-based STNO we make estimates for the critical current values. We use the following parameters of the system: $R_0=200$ nm, $R_i=150$ nm, $L=5$ nm, $h=0$, $\eta=5$, $\epsilon=1$. In typical magnetic materials which are used in spintronic devices (we use values for permalloy), the Gilbert damping is $\alpha\sim 0.01$, $M_s=8\cdot 10^5 \text{A} \text{m}^{-1}$, characteristic frequency  $\omega_m\approx 30$ GHz.
For these parameters we obtain $\omega_0=11.4$ GHz,  $J_{cr1}=0.055 \text{A}$, $J_{cr2}=0.0017 \text{A}$. Dividing by the current carrying area $S=5.5\cdot 10^{-10} \text{cm}^2$ we obtain the critical current densities $j_{cr1}=10^{8} \text{A} \text{cm}^{-2}$. The value of $j_{cr1}$ is high but reasonable \cite{Demidov}, especially having in mind that it should be applied for a short time only to trigger the oscillations. \as{The estimated time of its application is less than $0.1$ ns, which is short enough.} After the ignition the applied current can be reduced to smaller densities above the elimination threshold $j_{cr2}=3.1\cdot 10^{6} \text{A} \text{cm}^{-2}$, which is sufficiently low. This makes the ring-based STNO promising for the practical implementation. Let's consider the configuration of this device in more detail.

\section{Extracting an electric output signal \label{signal}}
In the previous sections we made theoretical description of the self-sustaining nonlinear dynamics of magnetization in the ferromagnetic ring. We suggest to use the ring as a free layer of the STNO. For its practical application as a microwave generator it is necessary to convert the energy of the oscillations into an alternating microwave electrical signal. There are two known phenomena allowing to extract electrical current from the magnetization oscillations: giant magnetoresistance (GMR)\cite{Katti,GMR} and the inverse spin Hall effect (ISHE)\cite{SH1,SH2}. 
In what follows we are considering both possibilities. 

\begin{figure}[h]
\includegraphics[width = 0.3\linewidth]{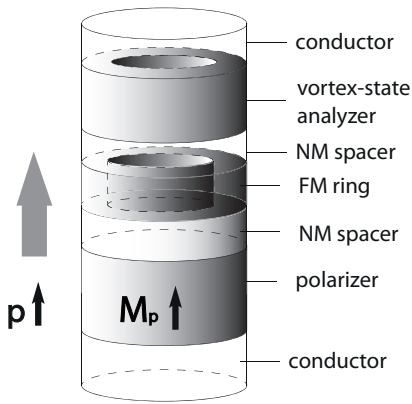}
\caption{Nanopillar geometry and the scheme of extracting the electric output signal by GMR. The analizer is a ferromagnetic ring (optionally a disk) in the vortex state. The gray arrow indicates the direction of the current flowing into the nanopiler (opposite to the direction of electron movement). Vector $\mathbf{p}$ shows the direction of spin polarization. \label{Fig3}}
\end{figure}

\begin{figure}[h]
\includegraphics[width = 0.3\linewidth]{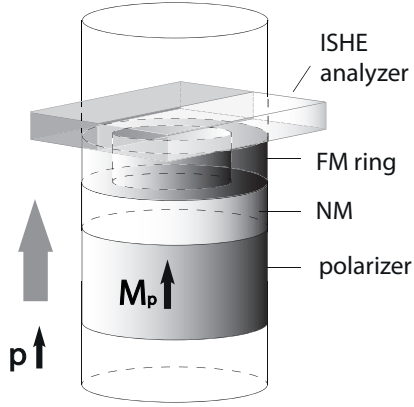}
\caption{Devise geometry and the scheme of extracting the electric output signal by ISHE. The nonhomogeneous analyzer contains two parts made of heavy metals with different signs of spin Hall angle. The gray arrow indicates the direction of the current flowing into the nanopillar (opposite to the direction of electron' motion). Vector $\mathbf{p}$ shows the direction of spin polarization.\label{Fig4}}
\end{figure}

Geometries of the STNO for these two cases shown in Figs.~\ref{Fig3},\ref{Fig4}.  The basic design is a sandwiched structure with three main elements: polarizer, free layer, and analyser. The direct current $J$ is injected in the device along its axis. Spin-polarization of electrons in the current, $\mathbf{p}$, is provided by a thick ferromagnetic polarizer of magnetization $\mathbf{M}_p$, which assumed to be fixed. The vector  $\mathbf{p}\| \mathbf{M}_p$ and directed perpendicular to the plane of the free layer. Polarizer and the free layer are separated by nonmagnetic spacer. As it is described above, the spin-polarized current flowing through the free layer leads to the nonlinear oscillations of its magnetization $\mathbf{M}$. The Analyzer, in turn, converts the oscillations into a useful signal.

In contrast to mono-domain oscillators, in the vortex-state free layer the oscillations of the total magnetic moment are negligibly small. The deviation of the magnetization from the ground state during the oscillations lies in the plane of the free layer, and its value averaged over the layer area equals to zero. For this reason, the question of the signal extraction is nontrivial. We suggest a solution of this problem by using special non-uniform analyzers.

First, we discuss the possibility to obtain an alternating electrical signal at the STNO output by the GMR effect. In the spin valve scheme based on GMR the analyzer consist of a ferromagnetic layer with the fixed magnetization $\mathbf{M}_a$ and a thin nonmagnetic interlayer separating it from the free layer of magnetization $\mathbf{M}$. The magnetization of analyser are usually fixed by usage of a hard ferromagnet of large thickness or by an additional antiferromagnetic layer. When the spin-polarized current passes through this system of layers, the scattering of electrons depends on the mutual orientation of the magnetizations of the layers. The GMR contribution to the resistance of the spin valve is locally (at each point of the layer area) proportional to the scalar product of these magnetizations $(\mathbf{M}\cdot\mathbf{M}_a)$. 

In the case under consideration the ground state of the free layer is the vortex state, and projection of its magnetization on a fixed direction averaged over the layer area is zero. Moreover, the same \as{is valid for the highly-nonlinear oscillational regime providing the generation}, see Fig.~\ref{Fig1}. Thus the use of a traditional homogeneous spin valve with $\mathbf{M}_a=const$ is impractical. 

Let consider an analyzer in a form of a thick magnetic ring (optionally a disk) with the saturation magnetization $M_{as}$ in a vortex state, which is stable enough even being in contact with free layer and under the action of the current. In order to stabilize the vortex ground state of the analyzer, it has to be made thick enough. Optionally, the analyzer in the vortex state can be made from a chiral magnet, where the vortex state is stabilized by the Dzyaloshinskii-Moriya interaction (DMI). 

The magnetization of this vortex analyzer can be written as $\mathbf{M}_a/M_{as}=-\mathbf{e}_x\sin(\chi + \varphi_0)+\mathbf{e}_y\cos(\chi + \varphi_0)$, where $\mathbf{e}_x$ and $\mathbf{e}_y$ are orthogonal unit vectors in the plane of the layer, and $\mathbf{e}_z$ is the unit vector along the device axis. Here the constant $\varphi_0 =0$ is for the ``standard'' vortices (Fig.~\ref{Fig1}(a)) with closed magnetic flux, and $\varphi_0 =\pm \pi/2$ for so-called radial vortices, stabilized by interfacial DMI. For the nonlinear regime of the oscillations the scalar product  
$(\mathbf{M}\cdot\mathbf{M}_a)=M_sM_{sa}\sin\theta_0\cos(\psi(t)+\varphi_0)$. So the GMR contribution to the electrical resistance $\Delta R(t)$ of this vortex spin valve scheme is proportional to $\sin\theta_0\cos [\psi(t)- \varphi_0]$. The factor $\sin\theta_0$ is determined by the external magnetic field, and the best efficiency of the vortex spin valve is achieved for zero external field, when $\theta_0=\pi/2$. 

We also consider an alternative possibility to extract a useful electric signal by the inverse spin Hall effect. Magnetization precession that appears in the ferromagnetic ring-shaped free layer can acts as a spin pump to an adjacent nonmagnetic metal. The direction of the spin current flow is perpendicular to the interface between the magnetic free layer and the nonmagnetic analyzer layer, which in our case coincides with $z-$axis. A spin current generated via the spin pumping mechanism, i.e. the direction of polarization of electrons in the current, $\boldsymbol{j}_{sp}\varpropto\mathbf{m}\times \partial \mathbf{m}/ \partial t$. It can be presented as a sum of in-plane and out-of-plane components,  $\boldsymbol{j}_{sp}=\boldsymbol{j}_{sp}^{in}+\boldsymbol{j}_{sp}^{out}$. The component $\boldsymbol{j}_{sp}^{in}$ lies in the plane of the layers and $\boldsymbol{j}_{sp}^{out}$ is along $\mathbf{e}_z$. The spin pumping generated due to the \as{radially-symmetric} mode is
\begin{eqnarray}\label{}
\boldsymbol{j}_{sp}\varpropto
\frac{1}{2}\sin 2\theta_0\left[\mathbf{e}_x\sin(\chi+\psi)-\mathbf{e}_y\cos(\chi+\psi)\right]\frac{\partial\psi}{\partial t}+\mathbf{e}_z\sin^2\theta_0 \frac{\partial\psi}{\partial t},
\end{eqnarray}
where $\psi=\psi(t)$ determines the spin precession. 

If the external magnetic field is absent the ground state of the ferromagnetic ring is the in-plane vortex with $\theta_0=\pi/2$ and only nonzero component of the spin pumping is  $\boldsymbol{j}_{sp}^{out}$, which, as we show below, does not lead to the generation of an electrical signal. However, in the presence of the magnetic field, i.e., for the cone state vortex, the situation is absolutely different.

This spin current injected into nonmagnetic metal conductor attached to the free layer can be converted into a charge current $\mathbf{j}_c$ via the inverse spin Hall effect
\begin{eqnarray}\label{}
\mathbf{j}_{c}\varpropto\theta_{SH}\boldsymbol{\mathbf{e}}_z\times \mathbf{j}_{sp},
\end{eqnarray}
where $\theta_{SH}$ is the spin Hall angle. It gives rise to an alternating electromotive force, the source of a useful microwave signal. In the considered geometry in-plane component of the spin current  $\boldsymbol{j}_{sp}^{in}$ is responsible for the occurrence of the electric current,

\begin{eqnarray}\label{jc}
\mathbf{j}_{c}\varpropto
\theta_{SH}\sin 2\theta_0\left[\mathbf{e}_x\cos(\chi+\psi) + \mathbf{e}_y \sin(\chi+\psi)\right]\frac{\partial\psi}{\partial t}.
\end{eqnarray}

Due to of the factor $\sin 2\theta_0$, the signal is nonzero only when external magnetic field is applied and the free layer is in the cone vortex state with $\theta_0<\pi/2$. The direction of the electric current $\mathbf{j}_{c}$ lies in the plane of the layers perpendicular to $\boldsymbol{j}_{sp}^{in}$. The maximum value of the $\sin 2\theta_0$ and hence the amplitude of the electrical signal is reached at $\theta_0=\pi/4$, that corresponds to the magnetic field $H\approx 2\sqrt{2}\pi M_s$. 

To extract the output signal by the inverse spin Hall effect the analyzer layer has to be made of the nonmagnetic metallic conductor. The signal that can be taken is proportional to the value of the current density Eq.~(\ref{jc})  averaged over the interface between the free layer and the analyzer layer, i.e $\int_{R_{i}}^{R_{o}}r dr\int_0^{2\pi}d\chi \;\mathbf{j}_{c}$. For the homogeneous  material of the analyzer this integral is equal to zero because of the alternating periodic functions $\sin(\chi+\psi)$ and $\cos(\chi+\psi)$. To solve this technical problem, it is advisable to fabricate different areas of the analyzer made of two heavy metals with different signs of spin Hall angle $\theta_{SH}$, e.g., platinum and tantalum, as has been realized in \cite{ISHE3}. Then the electrical signals from different areas are summed up. For example, we can select areas according to the sign of $\sin\chi$, then averaged in-plane component of the spin current is along $\mathbf{e}_x$ and the electrical current flows in the direction $\mathbf{e}_y$. 

In summary, the proposed ring-based STNO is compatible with both known mechanisms of microwave signal extraction: the GMR effect and ISHE. 

\section{Conclusions}
In the paper we investigated the nonlinear spin dynamics of the ferromagnetic ring in the vortex state under the action of the spin-polarized current and suggested the basic design of the ring-based vortex-state STNO. Proposed nanooscillator is characterised by the working frequencies up to $10$ GHz, that makes it advantageous over the disc-based vortex oscillators operating in the sub-GHz range.

\as{Indeed, the working frequency of STNO is determined by the first excited spin-wave mode. For the magnetic disk it is the gyromode, the collective mode, manifesting itself as a homogeneous precession of the vortex core. The frequency of the gyromode is very low, it is of the order of $1$ GHz. The high generation frequency of the ring-based STNO is achieved due to the fact that the ring-shape particle does not have the mode of core precession, because it does not have a vortex core region. The first mode exited by the spin-torque in the ring is the nonlinear breathing mode, that has higher frequency. It is the radially-symmetric in-plane oscillations of magnetization with small out-of-plane component. The regime of steady-state oscillations characterised by the full radially-symmetric rotation of the magnetization vector provides the generation of useful microwave signal. The frequency estimates for the permalloy ring of spatial dimensions usual for the oscillators give $11.4$ GHz.} 
 
 \as{Also it is worth noting, that due to the vortex state of the proposed STNO the in-plane total magnetic moment of the ferromagnet remains zero during the oscillations. Thus the working oscillator does not create stray fields and it is possible to pack the oscillators more densely, for example, for creating arrays of oscillators.
In the contrary, oscillator based on a homogeneous state or on a quasi-homogeneous state have nonzero magnetic moment in the ground state of the system. The rotation of this magnetic moment provides the generation, and such oscillators create disadvantageous stray fields. } 
 
We also found that the nonlinear spin oscillations are described by the simple and universal equation, Eq.~(\ref{diff_eq}). This equation appears for many physical problems; probably, the most important system is a point Josephson contact. This formal analogy allows usage of serious groundwork already developed for application of Josephson systems in electronics \cite{Josephson1, Josephson2}.  Despite the fact that the spin dynamics of considered device is described by the same equation as  in cases of Josephson contact and antiferromagnetic oscillator, the ring-based oscillator has different operational properties. Indeed, the useful alternating electrical signal obtained by the ring-based STNO has different behavior with increasing of the spin-polarized current, and, as a result, different mechanisms of signal extraction are needed. In the antiferromagnetic oscillator the alternating part of the signal is related to inhomogeneity of precession of the Neel vector and decreases with the current.  Due to the vortex state of the ring-based STNO the alternating signal can be obtained for the homogeneous precession of magnetization, and therefore there is no effect of signal decreasing. This fact simplifies the basic design of the analyzer for the ring-based STNO. Nevertheless, the signal extraction technology in the case has to be nontrivial, since the oscillation of magnetization averaged over the area of the ring is zero, which will result in a zero average electromotive force in output of standard schemes. We suggested to solve this problem by usage of special nonhomogeneous analyzers described in Sec.~\ref{signal}. We showed that the ring-based STNO can be used both in the spin valve circuit and in combination with the inverse spin Hall effect.

We expect the proposed device to have the following advantages. It can be made of convenient size: ring radii are of the order of hundreds of nanometers, and ring thickness is thin enough for spin transfer effects, namely $1-5 \text{nm}$. The system allows the easy adjustments of the main parameters, like working frequency, by application of a weak enough out-of-plane magnetic field.  It is also worth noting the linear growth of the generation frequency with the current.  The threshold current has usual values for the vortex STNOs: spin-transfer torque overcome the damping at current density about $10^{6} \text{A} \text{cm}^{-2}$, whereas the ignition current is of the order of $10^{8} \text{A} \text{cm}^{-2}$. In addition, it is possible to slightly reduce the elimination current by the magnetic field. 

Aside from the continuous auto-oscillations discussed in the paper, note that the system also allows some impulsive regimes, which can be used, for example, for creation of artificial neurons for neuromorphic computing, see Refs.\cite{inertia2022, Khymyn-neuron, Khymyn-logic}.  All of the above makes the proposed ring-based STNO interesting for practical applications.

\section{Acknowledgment}
V. Uzunova thank K. Byczuk for support and helpful discussions. The work of B.A. Ivanov was partly supported by National Scientific Foundation of Ukraine under Grant No. 2020.02/0261.


\begin{thebibliography}{99}

\bibitem{Slonczewski} K. Slonczewski, J. Appl. Phys. \textbf{45}, 375 (1974).

\bibitem{Kiselev}
S. I. Kiselev, J. C. Sankey, I. N. Krivorotov, N. C. Emley, R. J. Schoelkopf, R. A. Buhrman and D. C. Ralph, Nature \textbf{425},
380 (2003).

 \bibitem{Ralph} D. C. Ralph and M. D. Stiles, J. Magn. Magn. Mater. \textbf{320}, 1190 (2008).

\bibitem{Bader} S. D. Bader and S. S. P. Parkin, Spintronics, ed. by J. S. Langer, Ann. Rev. Condens. Matter Phys. \textbf{1}, 71 (2010).

\bibitem{BA7} A. Hirohata, K. Yamada, Y. Nakatani, I.-L. Prejbeanu, B. Dieny, P. Pirro, and B. Hillebrands, J. Magn. Magn. Mater., \textbf{509}, 166711 (2020).

\bibitem{Ruiz-Calaforra}
A. Ruiz-Calaforra, A. Purbawati, T. Br\"{a}cher, J. Hem, C. Murapaka, E. Jimenez, D. Mauri, A. Zeltser, J. A. Katine, M.-C. Cyrille, L. D. Buda-Prejbeanu, and U. Ebels,Appl. Phys. Lett. \textbf{111}, 082401 (2017).

\bibitem{ISHE1} V. E. Demidov, S. Urazhdin, H. Ulrichs, V. Tiberkevich, A. Slavin,
D. Baither, G. Schmitz, and S. O. Demokritov, Nat. Mater. \textbf{11}, 1028 (2012).

\bibitem{ISHE2} V. E. Demidov, S. Urazhdin, A. Zholud, A. V. Sadovnikov, S. O. Demokritov, Appl. Phys. Lett. \textbf{105}, 172410 (2014).

\bibitem{Houssameddine} D. Houssameddine, U. Ebels, B. Delaet, B. Rodmacq, I. Firastrau, F. Ponthenier, M. Brunet, C. Thirion, J.-P. Michel, L. Prejbeanu-Buda, M.-C. Cyrille, O. Redon and B. Dieny, Nature Materials \textbf{6}, 447 (2007).

\bibitem{Vaysset} A. Vaysset, C. Papusoi, L. D. Buda-Prejbeanu, S. Bandiera, M. Marins de Castro,
Y. Dahmane, J.-C. Toussaint, U. Ebels, S. Auffret, R. Sousa, L. Vila, and
B. Dieny1, Appl. Phys. Lett. \textbf{98}, 242511 (2011). \textbf{%
69}, 054429 (2004).

\bibitem{Murugesh} Ch. Sanid and S. Murugesh, Japan. J. Appl. Phys. \textbf{51}, No 6R (2012).

\bibitem{BA5} H. Jung, Y.-S. Choi, K.-S. Lee, D.-S. Han, Y.-S. Yu, M.-Y. Im, P. Fischer, and S.-K. Kim, ACS Nano, \textbf{6}, 3712 (2012).

\bibitem{GMR2} A. Slavin and V. Tiberkevich, IEEE Trans. Magn. \textbf{45}, 1875 (2009).

\bibitem{Soucaille} R. Soucaille, J.-V. Kim, T. Devolder, S. Petit-Watelot, M. Manfrini,
W. Van Roy and L. Lagae, J. Phys. D: Appl. Phys. \textbf{50} 085002 (2017).

\bibitem{Houshang} A. Houshang, R. Khymyn, H. Fulara, A. Gangwar, M. Haidar, S. R. Etesami, R. Ferreira, P. P. Freitas, M. Dvornik, R. K. Dumas and J. Akerman, Nature Communications \textbf{9}, No. 4374 (2018).

\bibitem{Consoloa} G. Consoloa, L. Lopez-Diaz, and L. Torres, Appl. Phys. Lett. \textbf{91}, 162506 (2007).

\bibitem{Tarequzzaman}
M. Tarequzzaman, T. Bohnert, M. Decker, J. D. Costa, J. Borme, B. Lacoste, E. Paz, A. S. Jenkins, S. Serrano-Guisan, C. H. Back, R. Ferreira and P. P. Freitas, Communications Physics \textbf{2}, No.
20 (2019).

\bibitem{Zeng}
Zh. Zeng, G. Finocchio, B. Zhang, P. Kh. Amiri, J. A. Katine, I. N. Krivorotov, Y. Huai, J. Langer, B. Azzerboni, K. L. Wang, and H. Jiang, Sci Rep. \textbf{3}, 1426 (2013).

\bibitem{GuslienkoAranda} K. Y. Guslienko, G. R. Aranda and J. Gonzalez,  J. Phys.: Conf. Ser. \textbf{292}, 012006 (2011).

\bibitem{Zeng1}
Zh. Zeng, G. Finocchio  and  H. Jiang, Nanoscale \textbf{5},
2219-2231(2013).

\bibitem{Zaspel_07} B. A. Ivanov and C. E. Zaspel, PRL  \textbf{99},247208 (2007).

\bibitem{Zvezdin2009} A. V. Khvalkovskiy, J. Grollier, A. Dussaux, Konstantin A. Zvezdin, and V. Cros, Phys. Rev. B \textbf{80}, 140401(R) (2009).

\bibitem{Prokopenko} O. V. Prokopenko, I. N. Krivorotov, E. N. Bankowski, T. J. Meitzler, V. S. Tiberkevich, A. N. Slavin, J. Appl. Phys. \textbf{114}, 173904 (2013).

\bibitem{ZaspelGalkinaIvanov19} C.E. Zaspel, E.G. Galkina, and B.A. Ivanov, Phys. Rev. Applied \textbf{12}, 044019 (2019).

\bibitem{Tserkovnyak} S. K. Kim and Y. Tserkovnyak, Appl. Phys. Lett. \textbf{111}, 032401 (2017).

\bibitem{BA3} G. Tong, Y. Liu, T. Cui, Y. Li, Y. Zhao, and J. Guan. Appl. Phys. Lett., \textbf{108}, 072905 (2016).

\bibitem{BA1} Y. Yang, X.-L. liu, J.-B. Yi, Y. Yang, H.-M. Fan, and J. Ding. J. Appl. Phys.,\textbf{ 111}, 044303  (2012).

\bibitem{BA2} X. L. Liu, Y. Yang, C. T. Ng, L .Y. Zhao, Y. Zhang, B. H. Bay, H. M. Fan, and J. Ding. Adv. Mater., \textbf{27}, 1939 (2015).

\bibitem{ZaspelJMM} C. E. Zaspel and B. A. Ivanov,  J. Magn. Magn. Mat. \textbf{286}, 366 (2005).

\bibitem{Oleg} D. J. Clarke, O. A. Tretiakov, G.-W. Chern, Ya. B. Bazaliy, and O. Tchernyshyov, Phys Rev B \textbf{78}, 134412 (2008). 

\bibitem{SCR} R. Khymyn, I. Lisenkov, V. Tyberkevych, B.A. Ivanov and A. Slavin, Sci. Rep. \textbf{7}, 43705 (2017).

\bibitem{Wysin_02} 
B. A. Ivanov and G. M. Wysin, Phys. Rev. B  \textbf{65},134434 (2002).

\bibitem{Zaspel05} B. A. Ivanov and C. E. Zaspel, Phys. Rev. Lett. \textbf{94}, 027205 (2005).

\bibitem{Zaspel_05} 
C. E. Zaspel, B. A. Ivanov, J. P. Park, and P. A. Crowell, Phys. Rev. B  \textbf{72}, 024427 (2005).

\bibitem{Guslienko} 
K. Yu. Guslienko, 
 Journal of Nanoscience and Nanotechnology \textbf{8}, 2745 (2008).

\bibitem{Otani}  
R. Antos, Y. Otani, and J. Shibata, Magnetic Vortex Dynamics, J. Physical Society of Japan \textbf{77}, 031004 (2008).

\bibitem{Sheka2004} D. D. Sheka, I. A. Yastremsky, B. A. Ivanov, G. M. Wysin and F. G. Mertens, Phys. Rev. B \textbf{69}, 054429 (2004).

\bibitem{ZaspelIvanov09} 
C. E. Zaspel, E. S. Wright, A. Yu. Galkin, and B. A. Ivanov, Phys. Rev. B  \textbf{80}, 094415 (2009).

\bibitem{Galkin} 
A. Yu. Galkin and B.A. Ivanov, J. Exp. Theor. Phys. \textbf{109}, 74 (2009).

\bibitem{Voronov} 
V. P. Voronov, B. A. Ivanov, and A. M. Kosevich, Zh. Eksp. Teor. Fiz. \textbf{84}, 2235 (1983).   

\bibitem{Mikeska1978} H.-J. Mikeska, 
J. Physics C-Solid State Physics \textbf{11}, L29-L32 (1978)

\bibitem{Kulagin}
B. A. Ivanov and N. E. Kulagin, 
J. Exp. Theor. Phys. \textbf{85}, 516 (1997).

\bibitem{inertia2022}
D. Markovic, M. W. Daniels, P. Sethi, A. D. Kent, M. D. Stiles, and J. Grollier, 
Phys. Phys. B \textbf{105}, 014411 (2022).

\bibitem{Kent}
A. Parthasarathy, E. Cogulu, A. D. Kent, and S. Rakheja, 
Phys. Rev. B  \textbf{103}, 024450 (2021).

\bibitem{Ivanov2020}
B.A. Ivanov, J. Exp. Theor. Phys. \textbf{131}, 95-112 (2020).

\bibitem{Josephson1}
A. Barone and G. Paterno, \textit{Physics and Applications of the Josephson Effect} (Wiley-VCH, Weinheim, 1982).

\bibitem{Josephson2}
S. Savel'ev, V. A. Yampol'skii, A. L. Rakhmanov, and F. Nori, Rep. Prog. Phys. \textbf{73}, 026501 (2010).

\bibitem{Demidov}
V. E. Demidov, S. Urazhdin, A.Zholud, A. V.Sadovnikov, S. O. Demokritov, Appl. Phys. Lett. \textbf{105}, 172410 (2014).

\bibitem{Katti} R. Katti, Proceedings of the IEEE \textbf{91}, no. 5, 687, (2003).

\bibitem{GMR}  E.Y. Tsymbal, D.G. Pettifor, Solid State Physics  \textbf{56},  113 (2001).
 
\bibitem{SH1}  B. F. Miao, S.Y. Huang, D. Qu, C. L. Chien,  Phys. Rev. Lett. \textbf{111}, 066602 (2013).

\bibitem{SH2} T. Kimura, Y. Otani, T. Sato, S. Takahashi, and S. Maekawa, Phys. Rev. Lett. \textbf{98}, 156601 (2007).

\bibitem{ISHE3} 
J. Li, C. B. Wilson, R. Cheng, M. Lohmann, M. Kavand,
W. Yuan, M. Aldosary, N. Agladze, P. Wei, M. S. Sherwin,
and J. Shi, 
Nature \textbf{578}, 70 (2020).

\bibitem{Lisenkov} I. Lisenkov, R. Khymyn, J. Akerman, N. X. Sun, B. A. Ivanov, \textbf{100}, Phys. Rev. B, 100409(R) (2019)

\bibitem{Khymyn-neuron} 
R. Khymyn, I. Lisenkov, J. Voorheis, O. Sulymenko, O.
Prokopenko, V. Tiberkevich, J. Akerman, and A. Slavin,
Sci. Rep. \textbf{8}, 1 (2018).

\bibitem{Khymyn-logic} 
O. Sulymenko,  O. Prokopenko, I. Lisenkov, J. Akerman, V. Tiberkevich, A. N. Slavin, and R. Khymyn, 
J. Appl. Phys.  \textbf{124}, 152115 (2018).

\end{thebibliography}
\end{document}